\documentclass[journal,twoside]{IEEEtran}
\usepackage{amsmath,amsfonts}
\usepackage{algorithmic}
\usepackage{algorithm}
\usepackage{array}
\usepackage[caption=false,font=normalsize,labelfont=sf,textfont=sf]{subfig}
\usepackage{textcomp}
\usepackage{stfloats}
\usepackage{url}
\usepackage{verbatim}
\usepackage{graphicx}
\usepackage{booktabs}
\usepackage[backend=biber]{biblatex}
\usepackage{hyperref}
\hypersetup{hidelinks,
	colorlinks=true,
	allcolors=black,
	pdfstartview=Fit,
	breaklinks=true}
\newtheorem{definition}{Definition}
\newtheorem{remark}{Remark}

\begin{document}

\title{Towards Emergency Scenarios: An Integrated Decision-making Framework of Multi-lane Platoon Reorganization}

\author{Aijing Kong, Chengkai Xu, Xian Wu, Xinbo Chen, Peng Hang,~\IEEEmembership{Senior Member,~IEEE} 
\thanks{This work was supported in part by the National Natural Science Foundation of China (52302502), the State Key Laboratory of Intelligent Green Vehicle and Mobility under Project No. KFZ2408, the State Key Lab of Intelligent Transportation System under Project No. 2024-A002, and the Fundamental Research Funds for the Central Universities.}
\thanks{Ajing Kong, Xian Wu and Xinbo Chen are with the school of Automotive Studies, Tongji University, Shanghai 201804, China. (e-mail: ajkong@tongji.edu.cn, wuxian@tongji.edu.cn, chenxinbo@tongji.edu.cn)}
\thanks{Chengkai Xu and Peng Hang are with the College of Transportation, Tongji University, Shanghai 201804, China. (e-mail: xck1270157991@gmail.com, hangpeng@tongji.edu.cn)}
\thanks{Corresponding author: Xian Wu, Peng Hang.}}


\maketitle

\begin{abstract} 
To enhance the ability for vehicle platoons to respond to emergency scenarios, a platoon distribution reorganization decision-making framework is proposed. This framework contains platoon distribution layer, vehicle cooperative decision-making layer and vehicle planning and control layer. Firstly, a reinforcement-learning-based platoon distribution model is presented, where a risk potential field is established to quantitatively assess driving risks, and a reward function tailored to the platoon reorganization process is constructed. Then, a coalition-game-based vehicle cooperative decision-making model is put forward, modeling the cooperative relationships among vehicles through dividing coalitions and generating the optimal decision results for each vehicle. Additionally, a novel graph-theory-based Platoon Disposition Index (PDI) is incorporated into the game reward function to measure the platoon’s distribution state during the reorganization process, in order to accelerating the reorganization process. Finally, the validation of the proposed framework is conducted in two high-risk scenarios under random traffic flows. The results show that, compared to the baseline models, the proposed method can significantly reduce the collision rate and improve driving efficiency. Moreover, the model with PDI can significantly decrease the platoon formation reorganization time and improve the reorganization efficiency.
\end{abstract}


\section{Introduction}
\IEEEPARstart{P}{latoons} of Intelligent and connected vehicles can potentially improve traffic efficiency by organizing vehicles in small spaces and improving fuel economy by reducing air resistance \cite{r35,pan2023research}. However, emergency environments challenge traditional single-lane vehicle platoons, which can only produce longitudinal acceleration/deceleration behaviors. Multilane vehicle platoons show superior capability in response to dangerous and complicated environments, through consideration of appropriate lane-changing and overtaking behaviors to achieve higher safety and efficiency. Currently, the main issue deserved special attention is how to make smart lane-changing and overtaking decisions for autonomous vehicle platoon in complex multilane environment \cite{xu2024towards}. 

An attempt to improve multilane vehicle platoons' safety and efficiency is the Platoon Cooperative Lane Change Control (P-CLCC) strategy, which can realize platoon lane-changing behaviors automatically. P-CLCC is designed based on two main control architectures: the multi-agent tracking control scheme and multi-agent motion planning structure. 

\romannumeral1) The multi-agent tracking control scheme contains the leader-follower architecture, virtual structure, consensus control system and so on. under this scheme, the leader vehicle drives referring to the reference trajectory, while other vehicles attempt to track the leader with a fixed formation \cite{li2023continual}, such as changing lanes simultaneously, changing lanes one by one and so on. However, the fixed-formation lane-changing behavior requires larger longitudinal space or longer moving time of and has poor ability to avoid dynamic obstacles such as a behind high-speed driving vehicle. Moreover, the rear-end collision is likely to happen as due to the untimely lane changing obstacle avoidance behavior. 

\romannumeral2) The multi-agent motion planning structure is dedicated to calculate feasible, safe, and comfortable trajectories for each member vehicle in platoon \cite{naeem2023optimal}. Modeling accurate trajectories of all vehicles through constrained centralized optimization can realize the platoon lane-changing behavior through flexible formations, and improve the safety efficiency and comfort simultaneously. However, the computational burden will increase rapidly with the platoon scale expands, which is deleterious to platoon decision-making process, especially in emergency scenarios with high-speed dynamic obstacles.

Furthermore, the P-CLCC system pays more attention to the motion planning process of vehicle platoons, most of which are engined by simple rule-based decision model. Complex environments challenge traditional P-CLCC systems. Hence, the group decision-making intelligence and cooperative capability of autonomous vehicle platoons need to be enhanced instantly. Inspired by cooperative Unmanned Aerial Vehicle (UAV) control system, the adaptive configuration transformation ability plays an important role for UAV group to respond to complex environments. Thus, we expect that the vehicle platoon can also posses the configuration transformation capability, which can change the platoon formation automatically to response to emergency environments and then reorganize to the original platoon configuration, in order to guarantee platoon safety and enhance traffic efficiency. Therefore, a platoon reorganization decision-making framework is presented.

In multi-agent systems, the decision-making and planning techniques of group reorganization process are generally game-based or Reinforcement-Learning-based (RL-based). Game-based strategies can make equilibrium decisions for each player by modeling the collaboration or competition relationships among agents. However, it's hard to directly build a clear game strategy during platoon reorganization process, in that the game objective dynamically changes as the platoon reconfiguration state evolves, avoiding emergency obstacles in early stage and gathering to original formation in late stage. While learn-based methods are well-versed in dealing with the issues of great variability and uncertainty, and can make proper decisions in real-time. 

Therefore, a Game/RL Dual-layer driven decision making and motion planning Framework (GRDF) for multilane autonomous vehicle platoons is proposed to realize the platoon reorganization process when facing emergency environments. With presented method, the vehicle driving safety will be enhanced, and the ability to cope with risks and traffic efficiency of vehicle platoons will be improved. The main contributions of this paper are summarized as follows:

\romannumeral1) An upper RL-based platoon behavior decision making model is presented, which can decide proper platoon formation according to the risk grade of different environments. A risk potential field model is established in the reward function to explicitly assess the risk of exploration actions, thereby improving the safety of the RL model. 

\romannumeral2) A lower coalition-game-based cooperative decision-making model is put forward, which is devoted to generating optimal behavior for each vehicle on the basis of current formation. A novel index based on the graph theory is proposed to measure the platoon formation distribution. Through verification, it has been shown that this index can significantly reduce the time required for platoon formation reorganization, improve reorganization efficiency, and thus enhance the overall traffic efficiency of the platoon.   

\romannumeral3) A dual-layer decision making and motion planning framework driven by RL and game is proposed for platoon formation reorganization. The method is verified in multiple typical emergency scenarios with uncertainty on the highway road, which shows better performance in safety and traffic efficiency.  

The remainder of this paper is as follows: Section \uppercase\expandafter{\romannumeral2} introduces related works. Section \uppercase\expandafter{\romannumeral3} establishes the overall framework of the platoon reorganization strategy. The method of making platoon-layer decisions and vehicle-layer decisions is represented in section \uppercase\expandafter{\romannumeral4} and \uppercase\expandafter{\romannumeral5}. Section \uppercase\expandafter{\romannumeral6} designs experiments involving multiple scenarios and thoroughly validates the effectiveness of the proposed strategy. Section \uppercase\expandafter{\romannumeral7} draws a clear conclusion of this paper.

\section{Related Works}
The related works are summarized from two aspects, the P-CLCC strategies for multi-lane vehicle platoons and the cooperative decision-making methods of intelligent vehicle platoons.

\subsection{P-CLCC Strategies}
The schemes of the P-CLCC methods can be divided into two main types.

\romannumeral1) Decoupling decision making and control for the vehicle platoon: This scheme usually adopts a rule-based model to make the lane change decisions of each vehicle in a certain order \cite{r17}. When the lane change space for the vehicle platoon is large enough, the vehicles are able to make lane change decisions together \cite{r1,r2}. When there is not enough free space, work \cite{r3} proposed a successive platoon lane change approach to improve the passing ability of the platoon in a high-density traffic environment. Moreover, works \cite{r4,r5,r6} put forward a particular P-CLCC logic that, the vehicle in platoon, who satisfies the safety conditions at the first place, should change to the target lane immediately, and then slow down to make enough space for other vehicles within a given time window. However, the predefined rules or logic for lane changing are not suitable for complex environments, especially when encountering aggressive human drivers or sudden emergency situations.

\romannumeral2) Coupling decision making and control for the vehicle platoon: This scheme allows vehicles in the platoon to make lane change decisions asynchronously, which can coordinate all platoon vehicles to a greater degree when faced with more complex and emergency environments. Most studies transform the cooperative lane change process into a Multiple Vehicle Motion Planning (MVMP) problem \cite{r7,r8,r9}. Work \cite{r8} proposed a MVMP method for platoon using Monte Carlo Tree Search, which splits the whole process into a series of lane change tasks and determines the optimal lane change sequence. The works \cite{r10,r11,r12} established a central Mixed Integer Linear Programming (MILP) model to calculate each vehicle's optimal discrete lane change decision. However, the MVMP model mainly focus on the trajectory planning process and it still needs clear inputs of starting and ending states of vehicles. Studies such as \cite{r13,r14,r15,r16} proposed multilayer decision structures, containing the upper platoon coordination layer and the lower vehicle motion planning layer, which can leverage the benefits of vehicle collaboration and improve model generalization. Works \cite{r13} and \cite{r14} presented a optimal-based model to generate the ideal formation order and motion planning was executed via Model Predictive Control (MPC) system. The dual-layer decision-making framework significantly outperforms the MVMP framework in terms of generalization and computational efficiency. However, pure optimization-based frameworks require the formulation of complex multi-variable and multi-objective coupled optimization problems, facing both solution difficulty and computational efficiency challenges.

\subsection{Cooperative Decision-making Methods of Multi-lane Platoons}
Existing Cooperative decision-making methods for autonomous vehicle platoons can generally be categorized into learning-based methods, optimization-based methods, and other approaches. Among the optimization-based methods, there are further distinctions between pure optimization and game-theoretic optimization approaches.

\romannumeral1) Learning-based methods: Longitudinal control of vehicle platoons using reinforcement learning or multi-agent reinforcement learning has become increasingly prevalent \cite{r18,r19,r20,r21}. Work \cite{r19} proposes a longitudinal control algorithm for platooning based on multi-agent reinforcement learning, which allocates reward functions according to the car-following relationships within the platoon to address the "lazy agent" problem. Work \cite{r21} introduces a multi-agent reinforcement learning algorithm aimed at enhancing safety in human-machine mixed platoons. The approach incorporates a CEF-QP constraint and an uncertainty-aware trajectory prediction model to improve control safety. However, these researches only account for uncertainties within the platoon, such as driver reaction times, while neglecting external uncertainties, such as obstacles ahead or sudden intrusions from other vehicles. Moreover, these learning-based methods have poor ability to efficiently coordinate the lateral maneuvers of the platoon, lack of adaptability to complex scenarios.

\romannumeral2) Optimization-based methods: Optimization-based and game theory-based methods are primarily applied to longitudinal and lateral cooperative control of vehicle platoons, collaborative obstacle avoidance decision-making, and decisions related to vehicle merging into or leaving the platoon \cite{r22,r24,r26}. Work \cite{r23} proposes a non-cooperative game decision-making model for platoons to handle scenarios involving lane-right conflicts. When an external vehicle attempts to cut into the platoon, the model determines the entry position of the vehicle within the platoon through a non-cooperative game, automatically adjusting the longitudinal gaps in the platoon. This method effectively mitigates collision risks caused by external vehicle cut-ins. However, the platoon passively accommodates the cut-in behavior, leading to disruption of its formation. Over time, this could result in reduced driving efficiency for the platoon. Work \cite{r25} introduces a game-theory-based cooperative obstacle avoidance method for vehicle platoons. This approach assumes that the platoon can evade obstacles and execute overtaking maneuvers by transitioning between predefined configurations. However, the method only considers a limited number of fixed configurations, and the transitions between these configurations are governed by simple logical rules. Consequently, the approach has limited adaptability to diverse and complex scenarios, as the transition rules lack sophistication and generalization capabilities.

\romannumeral3) Other methods: Work \cite{r28} proposes a rule-based and graph-theoretic approach for cooperative obstacle avoidance and decision-making in platoons. When obstacles are detected, the leading vehicle guides the following vehicles to perform formation adjustments by navigating through graph nodes to avoid the obstacle. Work \cite{r27,r29} introduces a multi-mode platoon decision control algorithm, dividing platoon control into functions such as cruising, following, lane-changing, overtaking, and parking. Mode transitions are managed via a state machine, aiming to enhance the adaptability of the platoon to diverse scenarios. However, most methods rely solely on physical rules tailored to specific scenarios, limiting their scalability and adaptability to dynamic environments. What's more, when external risks disrupt the platoon formation, these methods fails to provide effective countermeasures to restore the formation or mitigate the resulting safety risks.

\section{Overall Framework}
\subsection{Problem Statement}
Assume the traffic flow condition is a mixed traffic flow: the vehicle platoon consists of Connected and Autonomous Vehicles (CAVs), while the background traffic flow consists of Human-Driven Vehicles (HDVs). Within the platoon, vehicles share state information, which is transmitted following a specified communication topology. The traffic flow set is noted as $V$. 
\begin{equation}
\label{eq: 1}
    V = \left\{ V^c_1, V^c_2, V^c_3, V^h_4, V^h_5, ... \right\}
\end{equation}

\begin{figure}[!t]
\centering
\includegraphics[width=0.45\textwidth]{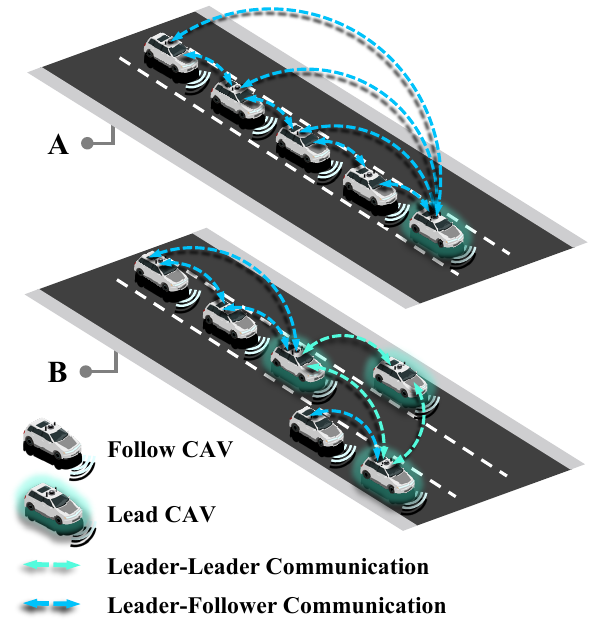}
\caption{The communication topology of vehicle platoon}
\label{fig:com}
\end{figure}

The platoon adopts a Leader-Leader-Predecessor-Follower (LLPF) communication topology, defined as follows: There is bidirectional communication between different platoon leaders. There is bidirectional communication between the leader and its directly connected followers within the platoon and between each follower and its immediate predecessor. The schematic diagram of the communication topology adopted by the platoon is shown in Figure \ref{fig:com}. The information exchanged between vehicles is represented in Equation \ref{eq: 2}. 

\begin{equation}
\label{eq: 2}
    F = \left\{ l_{target, i}, x_i, y_i, v_{x, i}, v_{y, i}, a_{x, i}, a_{y, i} \right\}
\end{equation}
where $l_{target, i}$ means the target lane number of the $i$-th vehicle, $x_i$, $y_i$, $v_{x, i}$, $v_{y, i}$, $a_{x, i}$ and $a_{y, i}$ denote the position, speed and acceleration information of the $i$-th vehicle.

The communication topology of the vehicle platoon is modeled on the basis of graph theory, noted as $G = (V^c, E)$. $E$ is the set of edges between vehicles. 

\begin{definition}[Platoon Communication Topology]
For $\forall i, \in , [0, n-1], \forall j, \in ,[1, n], E_{ij}$ satisfies: 
\begin{equation}
\label{eq: 3}
    E_{ij}=
\begin{cases}
1 & \mathrm{if} \, i=0 \; \mathrm{or} \, i<j \\
0 & \mathrm{otherwise}
\end{cases}
\end{equation}
\end{definition}

where $n$ represents the number of platoon vehicles, $0 \sim n$ means the numbering of all vehicles in the platoon in descending order based on their longitudinal positions. $E_{ij} = 1$ denotes the information can be transferred from $V^c_i$ to $V^c_j$.

To reduce additional computational complexity, the vehicles within the platoon are simplified using a kinematic model. 
\begin{align}
\label{eq: 4}
    \dot{y}\left(t\right)=u(t)\cos(\theta(t)) \\
    \dot{x}(t)=u(t)\sin(\theta(t)) \nonumber
\end{align}

\subsection{The Dual-layer Framework}

\begin{figure*}[!t]
\centering
\includegraphics[width=\textwidth]{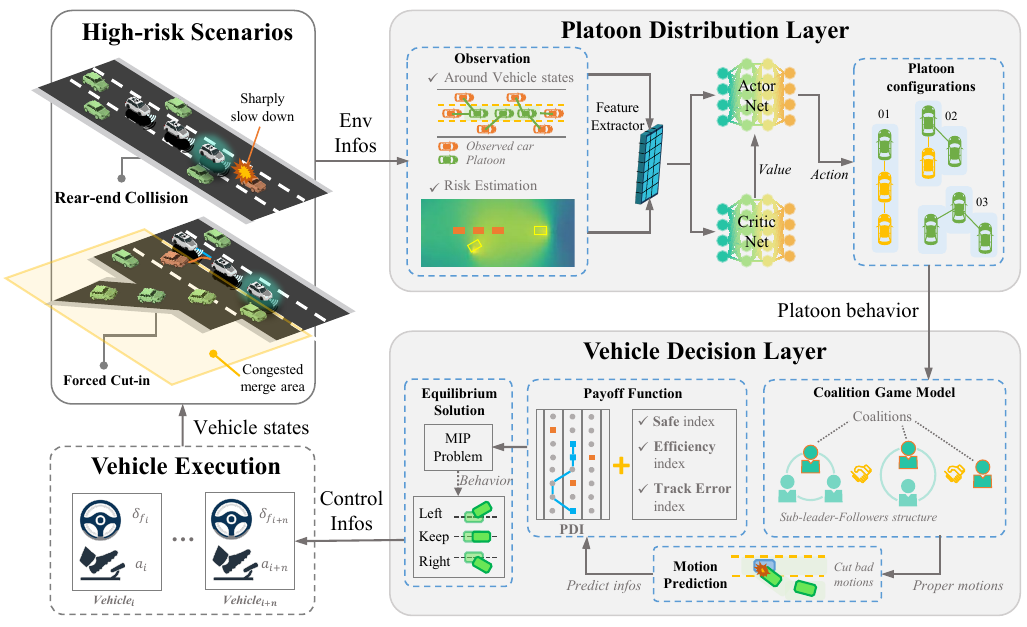}
\caption{The overall structure of GRDF framwork}
\label{fig: struct}
\end{figure*}

In this paper, we present a Game/RL Dual-layer driven decision making and motion planning Framework (GRDF) for multilane autonomous vehicle platoons in highway environments. The diagram of GRDF is shown in Figure \ref{fig: struct}. 

The GRDF framework proposed in this paper is divided into three main modules: the platoon distribution model, the vehicle cooperative decision-making model, and the vehicle execution model.

Environmental information including around vehicle states and risk, is input to the platoon distribution layer, and the output is the ideal platoon configuration. In this model, an RL-based decision-making method is proposed, which extracts necessary observation information from the dynamic environment and calculates the desired platoon grouping using an actor-critic network architecture. 

After receiving the platoon distribution behavior, the vehicle decision layer is supposed to generate the behavioral decision-making results for each vehicle in platoon. In this model, we propose a coalition game-based cooperative decision-making method and formulate an mixed-integer optimization problem (MIP) optimization problem to compute the game equilibrium solution. Additionally, to improve optimization efficiency, a pruning strategy is designed to eliminate actions that are infeasible or do not meet safety requirements. 

Finally, the vehicle execution model is created to transform the vehicle behavior into the longitudinal and lateral control information and make the vehicle drive in obedience to the given control commands. In this model, the motion planning is realized through lattice method.The trajectory clusters are generated using polynomials and a dynamics stability checker id designed to validate the trajectories. The best trajectory is selected by a comprehensive evaluation function, considering efficiency, safety, comfort and so on. The lower-level vehicle following and steering control are achieved using linear quadratic regulator (LQR) and Proportional Integral Derivative (PID) controllers.

\section{Platoon Distribution Layer}

\subsection{Preliminary Knowledge}
Reinforcement learning methods are based on Markov processes, where reward or penalty functions are predefined. Through training, agents interact with the environment to learn optimal strategies that maximize the reward or minimize the penalty function. Deep reinforcement learning further facilitates approximating highly nonlinear functions from complex datasets, making it well-suited for addressing complex decision-making and control problems \cite{r30}.

Early reinforcement learning algorithms, such as Deep Q Network (DQN), use neural networks to predict the Q-values of actions, aiming to minimize the discrepancy between the predicted Q-value network and the target Q-value network. The loss function can be defined as Equation \ref{eq: 5}. However, the process of optimizing and updating the action network may not align with the dynamic characteristics of the environment, presenting challenges for applications in complex and dynamic scenarios. 

\begin{equation}
    \label{eq: 5}
    L(\theta)=\mathbb{E}_{(s,a,r,s')}[(r+\gamma)^2]\mathop{\max_{a'}}Q_{\theta_{targt}}(s',a')
\end{equation}
where $(s,a,r,s')$ represents the transition tuple, $s$ is the current state, $a$ is the action taken, $r$ is the reward received, and $s'$ is the next state, $Q_\theta(s',a')$ is the Q-value estimated by the current network with parameters $\theta$, $Q_{\theta_{targt}}(s',a')$ means the target Q-value estimated by the target network with parameters $\theta_{target}$, $\gamma$ denotes the discount factor, which determines the weight of future rewards. 

The Proximal Policy Optimization (PPO) algorithm is implemented based on the actor-critic architecture. It approximates the optimal policy using the actor network and evaluates the policy's Q-values through the critic network, integrating policy gradient and value network update methods for more accurate policy evaluation and updates. Additionally, the PPO method introduces a clipping function to constrain the range of policy update gradients, thereby enhancing the stability of the learning process. The loss function for policy gradient updates is defined as Equation \ref{eq: 6}.

\begin{equation}
\label{eq: 6}
L^{FG}(\theta)=\mathbb{E}_t\left[\min\left(r_t(\theta)\hat{A}_t,\operatorname{clip}(r_t(\theta),1-\epsilon,1+\epsilon)\hat{A}_t\right)\right]
\end{equation}
where $r_{t}(\theta)=\frac{\pi_{\theta}(a_{t}|s_{t})}{\pi_{\theta_{\mathrm{old}}}(a_{t}|s_{t})}$ is the probability ratio between the new policy $\pi_\theta$ and the old policy $\pi_{\theta_{old}}$, $\hat{A}_t$ represents the advantage estimate at time step $t$, and $\epsilon$ is a hyperparameter controlling the clipping range to balance exploration and stability during policy updates.

\subsection{Centralized Proximal Policy Optimization Algorithm}
In this section, we propose a platoon configuration decision-making method based on the PPO algorithm. The leading vehicle can obtain the state and perception information of other vehicles through the communication topology, enabling information sharing within the platoon. Based on this information-sharing mechanism, the leading vehicle performs centralized decision-making for platoon configuration. 

The decision-making process is modeled as a general Partial Observation Markov Decision Process (POMDP), $(S,A,P,R,\gamma,O,Z)$. Among these elements, the action space $A$, observation space $O$, and reward function $R$ are particularly important. 

\begin{figure}
\centering
\includegraphics[width=0.4\textwidth]{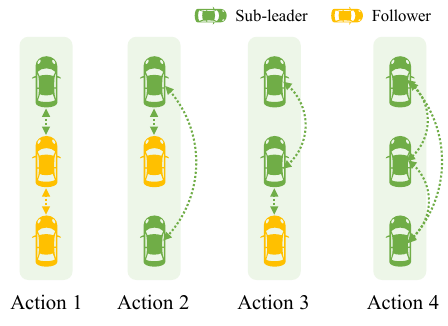}
\caption{The platoon configuration actions}
\label{fig: rl action}
\end{figure}

The action space $A$ includes various platoon configurations, defined in Equation \ref{eq: 7}. Taking 3-vehicle platoon as an example, the platoon configuration action is illustrated in Figure \ref{fig: rl action}. 

\begin{align}
\label{eq: 7}
A &= \{ a_1, a_2, a_3, \ldots, a_m \}  \\
a_1 &= [(0,1,2,\ldots,m)] \nonumber \\
a_2 &= [(0)(1,2,\ldots,m)] \nonumber \\
a_3 &= [(0)(1)(2,3,\ldots,m)] \nonumber
\end{align}

where $a_i$ represents each platoon configuration. $m$ is the total number of a $n$-vehicle platoon's configuration, $m=C^1_{n-1}+C^2_{n-1}+...+C^{n-1}_{n-1}+1$. 

The observation space $O$ denotes the observed environment information accessible to the controlled platoon, including positions, speeds, accelerations and jerk of objects. Additionally, we incorporated the TTC index into the observation space as an explicit representation of environmental risk states. This enhancement improves the algorithm's adaptability to high-risk scenarios by providing a quantitative measure of potential collision risks. The observations are obtained through sensors and communication units with random noise. The observation space $O$ is defined in Equation \ref{eq: 8}.

\begin{equation}
    \label{eq: 8}
    O=\{ x,y,v_x,v_y,a_x,a_y,j_x,j_y,\tau\}
\end{equation}
where $x,y_x,v_y,a_x,a_y,j_x,j_y$ represent longitudinal and lateral position, speed, acceleration and jerk of observed objects. The values of $a_x,a_y,j_x,j_y$ are derived by $v_x,v_y$. $\tau$ means the TTC between ego vehicle and observed objects. 

The reward function $R$ guides the training process by defining the awards and penalties for decision-making results. The reward function primarily considers safety $R_s$, efficiency $R_e$, platoon driving performance $R_d$ and platoon reorganization performance $R_r$. 

\begin{equation}
    \label{eq: 9}
    R=w_s \times R_s+w_e \times R_e+w_d \times R_d+w_r \times R_r
\end{equation}

The safety reward $R_s$ consists of the collision penalty $r_{col}$ and the risk penalty $r_{ris}$. 

\begin{equation}
\label{eq: 10}
    R_s=w_{col} \times r_{col} + w_{ris} \times r_{ris}
\end{equation}

\begin{equation}
\label{eq:11}
r_{col} = \begin{cases}
0, & \text{if collision happened} \\
1, & \text{else}
\end{cases}
\end{equation}

\begin{figure}
\centering
\includegraphics[width=0.5\textwidth]{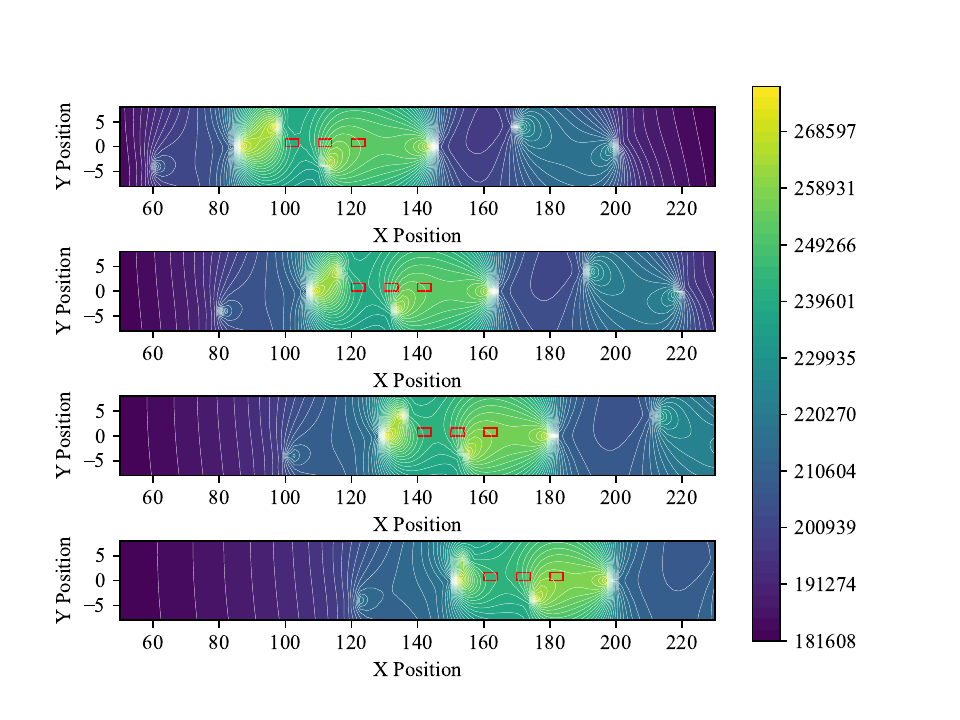}
\caption{The risk field distribution}
\label{fig: risk field}
\end{figure}

The risk reward $r_{ris}$ is characterized by the field intensity of the driving risk field, determined in Equation \ref{eq: 12}. The detailed construction of the risk field refers to work \cite{r31}, the scenario risk distribution is shown in Figure \ref{fig: risk field}.

\begin{equation}
    \label{eq: 12}
    r_{ris} = \mathop{\max_{i \in [1,n]}} (\frac{GRM}{d_i^{k_1}})^{k_2 v_i} / (\frac{GRM}{d_{\min}^{k_1}})^{k_2 v_{\max}}
\end{equation}

The efficiency reward $R_e$ is the platoon's average speed. 

\begin{equation}
\label{eq: 13}
    R_e = \frac{1}{n} \sum_{i=1}^n \frac{v_i}{v_{\max}}
\end{equation}

The platoon driving performance reward $R_d$ contains the tracking error of distance, lateral position and speed. 

\begin{align}
\label{eq: 14}
    R_d = \frac{1}{n-1} \sum_{i=1}^{n-1} & w_x \times (x_i - x_{i+1} - d_{target}) \; + \\
    & w_y \times (y_i - y_{i+1}) \; + \nonumber \\
    & w_v \times (v_i - v_{i+1}) \nonumber
\end{align}

where the target follow distance $d_{target}$ in platoon is 10 m.

The platoon reorganization performance reward $R_r$ is composed of platoon reorganization frequency $r_{rf}$, reorganization time $r_{re}$ and reorganization incentive $r_{re}$, shown in Equation \ref{eq: 15}

\begin{equation}
\label{eq: 15}
    R_r = w_{rf} \times r_{rf} + w_{re} \times r_{re} + w_{ri} \times r_{ri} 
\end{equation}

The platoon reorganization frequency reward $r_{rf}$ means the number of platoon reorganization events triggered per unit time, which is defined and normalized in Equation \ref{eq: 16}. 

\begin{equation}
    \label{eq: 16}
    r_{rf} = \frac{N_{trigger}}{N_{step}}
\end{equation}
where $N_{trigger}$ refers to the decision steps involved in switching the platoon from a single sub-platoon configuration to other multiple sub-platoon configurations during a single episode. $N_{step}$ represents the total decision steps within a single episode. 

The platoon reorganization time $r_{re}$ represents the average time taken by the platoon to successfully complete a reorganization process. Successfully completing a reorganization process means that the platoon switches from a single sub-platoon configuration to a multi-sub-platoon configuration and then back to a single sub-platoon configuration. 

The platoon reorganization incentive $r_{rc}$ denotes the degree of incentive for triggering platoon reorganization behavior under varying environments. When the platoon encounters a sudden decrease in environmental risk, denoted by TTC or a significant decline in efficiency, denoted by $R_e$, higher reorganization incentive is provided. Conversely, when $r_{ris}$ remains stable and the $R_e$ keeps in a higher level, a lower reorganization incentive is applied. 

\begin{align}
\label{eq: 17}
r_{rc} = k_t \times \frac{\tau_{\min}^0}{\tau^0} + k_v \times \sum_{i=1}^n \frac{v_i}{v_{\max}}
\end{align}

where $\tau^0$ represents the TTC between platoon leader and front object, $\tau_{\min}^0$ is the critical value of TTC, set as 2.5 s. $k_t$, $k_v$ are constant parameters. 

\section{Vehicle Decision Layer}
After the platoon-layer decision-making model calculates the optimal platoon configuration, this section establishes a coalition game model to determine the behavioral decisions for each vehicle.

\subsection{Formulation of coalition game decision-making model}
Coalition game is a type of cooperative game, which usually involves multiple players, denoted by $\mathcal{N}=\{ 1,...n \}$. who seek to form cooperative groups to achieve the optimal collective payoff distribution. Each group is called a coalition. Players within a coalition reach an agreement and act as a unified entity in the collaborative decision-making process. $v$ represents the quantified benefit that the coalition achieves during the game. The definition of $\mathcal{v}$ determines the specific form and type of the game. Hence, the coalition game can be described by a pair $(\mathcal{N}, v)$. 

The characteristic function $v$ can be defined in two ways \cite{r32}. In coalitional games with nontransferable utility (NTU), the payoff of each coalition $S$ is closely related to the joint actions of other coalitions. In games with transferable utility (TU), however, the payoff of each coalition is independent of the performance of other coalitions.

\begin{definition}[The characteristic function $v$ in NTU games]
In a NTU game $G = (\mathcal{N}, v)$, coalition $S$ is a subset of $\mathcal{N}$. All coalitions other than $S$ are supposed to form a complementary coalition, denoted as $\mathcal{N} \backslash S$. Treat $S$ and $\mathcal{N} \backslash S$ as two players participating in a non-cooperative game. The actions of coalition is denoted by $a$, and the profit of actions is denoted by $u$. The characteristic function $v$ of NTU game can be calculated by the maximin principle in equation \ref{eq: 18}, which means that the sum of the maximum payoffs guaranteed to coalition $S$ members when facing the most severe threats from coalition $\mathcal{N} \backslash S$ is ensured.

\end{definition}

\begin{equation}
    \label{eq: 18}
    v(S) = \mathop{\min_{\mathcal{N} \backslash S}} \mathop{\max_S} \sum_{i \in S} u_i(a_S,a_{\mathcal{N} \backslash S})
\end{equation}
where $a_S$ and $a_{\mathcal{N} \backslash S}$ means the actions of coalition $S$ and complementary coalition $\mathcal{N} \backslash S$. Natural convention $V(\emptyset)=0$. 

\begin{definition}[The characteristic function $v$ in TU games]
In a TU game $G = (\mathcal{N}, v)$, coalition $S$ is a subset of $\mathcal{N}$. The actions of coalition is denoted by $a$, and the profit of actions is denoted by $u$. The characteristic function $v$ of TU game can be defined in Equation \ref{eq: 19}.

\begin{equation}
    \label{eq: 19}
    v(S) = \mathop{\max_S} \sum_{i \in S} u_i(a_S, a_{\mathcal{N} \backslash S})
\end{equation}

\end{definition}

The NTU games contains implicit competitive relationship among coalitions. While in TU games, coalitions are in a totally cooperative relationship. In this paper, all vehicles within the platoon are committed to cooperation, thus a TU game is established.

In our platoon reorganization problem, the coalition $S$ needs to satisfy follow conditions.

\begin{remark}[Vehicle coalition formation conditions]
When the vehicles drive on the same lane and no other obstacles exist between them, a coalition can be formed from these vehicles. The positions of these vehicles are limited by Equation \ref{eq: 20}. Each coalition is a sub-platoon. The vehicle at the forefront becomes the leader of the sub-platoon. Based on the LLPF communication topology, the vehicles within the coalition and the sub-leaders between coalitions can exchange driving information.

\begin{align}
\label{eq: 20}
    \mathop{\max_{i} |x_i-x_{i+1}|} < \widetilde{x}_{lim}  \\
    \mathop{\max_{i} |y_i-y_{i+1}|} <  \widetilde{y}_{lim} \nonumber
\end{align}

where $\widetilde{x}_{lim}$ and $y_{lim}$ are constants, $\widetilde{x}_{lim} = 30$ m and $y_{lim} = 1.5$ m.
\end{remark}

\subsection{Coalition characteristic function}
The goal of platoon decision-making is to ensure that each sub-platoon, or coalition, completes the configuration transformation process safely and efficiently. This process can be divided into two phases: platoon splitting and platoon merging.

During the platoon splitting phase, each coalition strives to make full use of road space for coordinated lane changes to enhance the safety and efficiency of the grand coalition. In the platoon merging phase, each coalition aims to form a new platoon configuration as quickly as possible while ensuring its own safety and maintaining driving efficiency.

The characteristic function $v_p(S)$ of platoon splitting and platoon merging process is shown in Equation \ref{eq: 21}. 

\begin{align}
\label{eq: 21}
v_p(S) & = \mathop{\max_S} \sum_{i \in S} J_{sp}^i (\alpha_i, \beta_i) \\
v_p(S) & = \mathop{\max_S} \sum_{i \in S} J_{mg}^i (\alpha_i, \beta_i) \nonumber
\end{align}

where $J_{sp}(\alpha_i, \beta_i)$ and $J_{mg}(\alpha_i, \beta_i)$ are the profit of platoon splitting and platoon merging process. $\alpha_i$ and $\beta_i$ are the $i$-th action in $a_S$ and $a_{\mathcal{N} \backslash S}$, the action list of current coalition $S$ and the complementary coalition $\mathcal{N} \backslash S$. The action $\alpha$, $\beta$ refers to the lateral decision of each vehicle, $\alpha, \beta \in \{ left \; lane \; change, \; right \; lane \; change \;, keep \; the \; lane \}$ 

The splitting profit $J_{sp}(\alpha_i, \beta_i)$ includes safety profit $J_s(\alpha_i, \beta_i)$, driving efficiency profit $J_e(\alpha_i, \beta_i)$ and platoon integration profit $J_{it}(\alpha_i, \beta_i)$. The merging profit $J_{mg}\alpha_i, \beta_i)$ contains formation error $J_{tr}(\alpha_i, \beta_i)$ on the basis of the splitting profit $J_{sp}(\alpha_i, \beta_i)$. $J_{sp}(\alpha_i, \beta_i)$ and $J_{mg}(\alpha_i, \beta_i)$ are calculated by Equation \ref{eq: 22}. 

\begin{align}
\label{eq: 22}
J_{sp}(\alpha_i, \beta_i) = & w_s \times J_s(\alpha_i, \beta_i) + w_e \times J_e(\alpha_i, \beta_i) \\
 & + w_{it} \times J_{it}(\alpha_i, \beta_i) \nonumber \\
J_{mg}(\alpha_i, \beta_i) = & w_s \times J_s(\alpha_i, \beta_i) + w_e \times J_e(\alpha_i, \beta_i) \nonumber \\
       & + w_{it} \times J_{it}(\alpha_i, \beta_i) \nonumber \\
       & + w_{er} \times J_{er}(\alpha_i, \beta_i) \nonumber
\end{align}

where $w_s$, $w_e$, $w_{it}$ and $w_{er}$ are constant weights.


The safety profit $J_s(\alpha_i, \beta_i)$ is calculated in Equation \ref{eq: 23}.

\begin{align}
\label{eq: 23}
J_s(\alpha_i, \beta_i) = & (P_{ris}(\alpha_i, \beta_i) + k_{\tau} \times \tau^0(\alpha_i, \beta_i) \\
& + k_d \times ((x_{i} - x^{LV}_{i})^2 + (y_{i} - y^{LV}_i)^2)) \nonumber
\end{align}

where $P_{ris}(\alpha_i, \beta_i)$ denotes the value of risk field, which can be calculated in Equation \ref{eq: 12}. $(x_{i}$ and $y_{i}$ are the predicted longitudinal and lateral positions of vehicle $i$. $LV$ refers to the near front vehicle of vehicle $i$. $x^{LV}_{i}$ and $y^{LV}_{i}$ denote the position of $LV$. $\tau^0(\alpha_i, \beta_i)$ means the TTC value between vehicle $i$ and $LV$. $k_\tau$ and $k_d$ are constant values. 

The efficiency profit $J_e(\alpha_i, \beta_i)$ is derivated in Equation \ref{eq: 24}.

\begin{align}
\label{eq: 24}
J_e(\alpha_i, \beta_i) = v^{x,ave}_{i}
\end{align}

where $v^{x,ave}_{i}$ represents the average longitudinal speed in prediction time of vehicle $i$. 

The platoon integration profit $J_it(\alpha_i, \beta_i)$ refers to the performance of maintaining the integrity of the platoon formation during the platoon reorganization process, which can be represented by traffic entropy, shown in Equation \ref{eq: 25}.

\begin{equation}
    \label{eq: 25}
    J_{it}(\alpha_i, \beta_i) = -n_S \sum_{j=1}^l (\frac{p_j}{n_S} \ln(\frac{p_j}{n_S}))
\end{equation}
where $n_S$ is the number of vehicles in coalition $S$, $l$ denotes the number of lanes in roads, $p_j$ is the number of the $j$-th lane, including both ego vehicles and background vehicles. 

The tracking profit $J_{tr}$ is characterized by the position and speed error of vehicles within ego platoon, denoted in Equation \ref{eq: 26}.

\begin{align}
\label{eq: 26}
J_{tr}(\alpha_i, \beta_i) = \sum_{j=1}^{n_S-1} & ((x_j - x_{j+1} - d_{target}) + \\
& k_y (y_j - y_{j+1}) + k_v (v_j - v_{j+1})) \nonumber
\end{align}

where $x_j$, $y_j$, $v_j$ are the lateral position, longitudinal position, and speed of the $j$-th vehicle in the platoon, $x_{j+1}$, $y_{j+1}$, $v_{j+1}$ are the lateral position, longitudinal position, and speed of the $(j+1)$-th vehicle in the platoon

\subsection{Platoon Disposition Index (PDI)}
To further improve the performance of the platoon reorganization process, decrease the reorganization time, and enhance reorganization efficiency, we propose the Platoon Disposition Index (PDI) based on graph theory. 

The PDI serves to comprehensively assess the environmental situation, providing a real-time representation of the spatial distribution of vehicles within the platoon, as well as the difficulty of transitioning from the current distribution to the ideal one. The PDI not only considers the positional information of the vehicles within the platoon but also takes into account the distribution of surrounding vehicles in the environment. In contrast, common error metrics such as longitudinal or lateral position error only indirectly reflect the distribution of platoon vehicles through error terms, without considering the distribution of other vehicles in the environment. 

\begin{figure}[!t]
\centering
\includegraphics[width=0.45\textwidth]{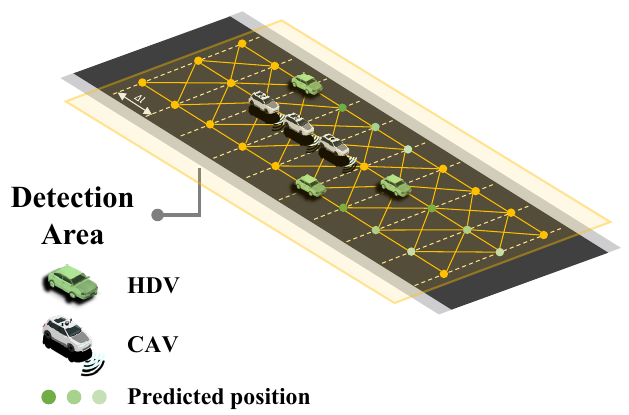}
\caption{Diagram of road node division}
\label{fig: node}
\end{figure}

The PDI is calculated based on graph theory. Thus, we simplifies the road into a node map, where each vehicle or obstacle occupies a node. The schematic diagram of road node division is shown in Figure \ref{fig: node}. $d_n$ is the distance between nodes, $d_n \in [d_n^{\min},d_n^{\max})$. The values of $d_n^{\min}$ and $d_n^{\max}$ are related to the vehicle's dimensions. In this paper, $d_n^{\min} = 10$ m and $d_n^{\max} = 20$ m. 

Before calculating PDI, The following assumptions are made for the nodes:

\romannumeral1) The node occupied by the first vehicle of the platoon is considered the starting point.

\romannumeral2) The node occupied by the last vehicle of the platoon is considered the end point.

\romannumeral3) Nodes occupied by background vehicles or obstacles are considered infeasible nodes. 

\romannumeral4) Only adjacent nodes can be connected to form an edge. the adjacent nodes are defined in Definition \ref{adjacent nodes}.

\romannumeral5) The weight of an edge is defined by both the distance between the two nodes and the lanes in which the nodes are located. Its physical meaning is the difficulty for a vehicle to move from one node to another, named Equivalence Distance (ED), is defined in Definition \ref{edge weight}. 

\begin{definition}[Adjacent Nodes]
\label{adjacent nodes}
    Node $n_i \in N_{n}$ and node $n_j \in N_{n}$. $N$ means the set of all nodes of the map. $L_i$ and $L_j$ are the lanes that $n_i$ and $n_j$ belong to. $x_i$ and $x_j$ are the longitudinal positions of $n_i$ and $n_j$. If a vehicle occupies a node, the node is located at the center of the vehicle's rear axle. If a node is occupied by an obstacle, the node is located at the center of the physical outline of the obstacle. If $L_i = L_j$ and $|x_i - x_j| <d_n^{\max}$, then $n_i$ and $n_j$ are adjacent nodes. If $|L_i - L_j| = 1$ and there exits a $n_k$ such that $|x_i-x_j| \leq |x_k-x_i|$, then $n_i$ and $n_j$ are adjacent nodes. 
\end{definition}

\begin{definition}[Equivalence Distance]
    \label{edge weight}
    $E_{ij}$ denotes the edge between node $n_i$ and node $n_j$. The $ED_{ij}$ is the weight of $E_{ij}$. The value of $w_{ij}$ id given by Equation \ref{eq: 27}.

    \begin{equation}
    \label{eq: 27}
        E_{ij} = \frac{\tilde{d}_{ij}}{D} + k_l |L_i - L_j| 
    \end{equation}
where $\tilde{d}_{ij}$ refers to the Euclidean distance between $n_i$ and $n_j$, $k_l$ is a constant value. In this paper, $k_l = 10$. 
    
\end{definition}

\begin{figure}
\centering
\includegraphics[width=0.5\textwidth]{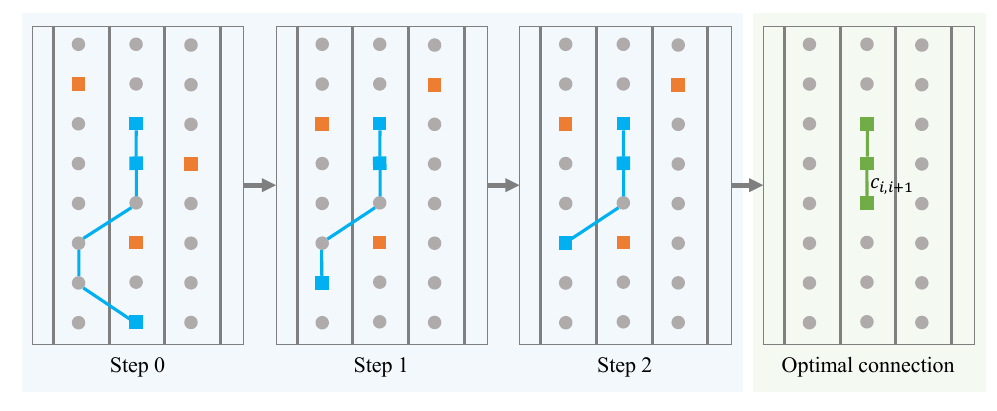}
\caption{PDI calculation process}
\label{fig: PDI}
\end{figure}

Based on the above assumptions \romannumeral1) $\sim$ \romannumeral5), the goal is to find the shortest path in terms of equivalent distance from the starting point to the endpoint, with the constraint that the path cannot pass through nodes occupied by background vehicles or obstacles. The sum of the ED between all the nodes along this path is defined as the equivalent length of the path. We will use the equivalent length of the path as the PDI. The schematic of the PDI calculation process is shown in Figure \ref{fig: PDI}.

The process of finding the shortest path is modeled as a MIP to solve for the PDI. All edges in the graph are set as 0-1 integer variables, denoted by $u$. The number of variables is $n_u$ The objective function is given by Equation \ref{eq: 28}. The constraints are shown in Equation \ref{eq: 29}.

\begin{equation}
    \label{eq: 28}
    J_{path} = \sum_{i=1}^{n_u} ED_i u_i
\end{equation}

\begin{equation}
    \label{eq: 29}
    \left \{
    \begin{array}{l}
    \sum_{j=1}^{n_u} u_{ij} = 1, \quad if \; n_i = n_0 \\
    \sum_{i=1}^{n_u} u_{ij} = 0, \quad if \; n_j = n_0 \\
    \sum_{i=1}^{n_u} u_{ij} = 1, \quad if \; n_i = n_e \\
    \sum_{j=1}^{n_u} u_{ij} = 0, \quad if \; n_j = n_e \\
    \sum_{i=1}^{n_u} u_{ij} = \sum_{i=1}^{n_u} u_{ji}, \quad if \; n_j \neq n_0 \; and \; n_j \neq n_0
    \end{array}
    \right.
\end{equation}

The PDI can be computed using common MIP solving methods, such as branch and bound, etc. The advantages of PDI mainly lie in: When vehicles in the platoon are in the same position but the positions of surrounding vehicles differ, traditional error metrics cannot distinguish such situations. For example, when the platoon is cut in by other vehicles, the traditional error indexes will make conservative decisions and guide the platoon to drive behind the cut-in vehicle, maintaining its original lane, in order to minimize tracking error indexes. However, by adopting PDI, the optimal decision is to change lanes to the left, allowing the platoon to transform into a tighter configuration more quickly, thereby improving platoon efficiency and fuel economy.

\section{Simulation and Discussion}

\subsection{Simulation Settings}
The simulation platform is built on the basis of an OpenAI Gym environment. In the simulator, actions determined by specific policies are translated into low-level steering and acceleration signals through a closed-loop PID controller. The vehicles in the simulator are built on the basis of kinematic models. The longitude and lateral decisions of HDVs are controlled by Intelligent Driver Model (IDM) and Minimizing Overall Braking Induced by Lane Change (MOBIL) models, respectively. 

In our simulation, the vehicles in the background traffic flow are generated randomly, with different driving styles to simulate the uncertainty of the environment. Due to communication distance and latency constraints, the number of vehicles in a platoon is typically limited to 2 to 5. In our case, the number of platoon vehicles is 3. All vehicles in the platoon are equipped with advanced autonomous driving capabilities and V2X or DSRC communication technologies. 

In the proposed GRDF model, we adopt the Multi-layer Perception (MLP) as the encoder and decoder, which contains one hidden layer and two fully connected (FC) layers. The size of hidden and FC layers is 256. The detailed hyperparameter is shown in Table \ref{table DPL}

\begin{table}
\begin{center}
\caption{Hyperparameters of DPL Model}
\label{table DPL}
\begin{tabular}{c c c}
\toprule
Symbol & Definition & Value \\
\midrule
$N_{t}$ & Total Training Steps & $10^{6}$\\
$\lambda$ & Learning Rate & $3^{-4}$ \\
$\gamma$ & Discount Factor & $0.85$ \\
$B$ & Batch Size & $64$ \\
$\epsilon$ & Clip Parameter & $0.2$ \\
$S_{u}$ & Number of Forward Steps & $128$ \\
$\tau$ & Target Update Rate & $0.0$ \\
$c_{v}$ & Value Loss Coefficient & $0.5$ \\
$c_{e}$ & Entropy Term Coefficient & $0.0$ \\
\bottomrule
\end{tabular}
\end{center}
\end{table}

\subsection{Case Studies}
When driving in highway roads, the platoon faces risks mainly from two directions, longitudinal and lateral. The longitudinal risk means potential collisions coming from the same lane as platoon. For instance, the front vehicle suddenly decelerates and causes a rear-end collision, which occurs frequently in high-speed traffic. The lateral risk indicates possible collisions caused by lateral adjacent vehicles as some drivers may take aggressive lane-change or overtake behaviors in order to pursue higher speed, which may lead to corner collision unluckily. Especially in ramp merging area, continuously vehicles merge into the merging lane of the main road, which brings about temporary congestion in merging lane. When platoon drives in uninterrupted near lane, the vehicles in merging lane is likely to cut into the platoon, breaking platoon formation and increasing lateral collision risk.

Hence, in this paper, we organize two simulation cases of longitudinal and lateral driving risks to evaluate the performance of proposed model, respectively. 

\subsubsection{Case 1: Lateral Driving Risk}

\begin{figure*}[!t]
\centering
\includegraphics[width=\textwidth]{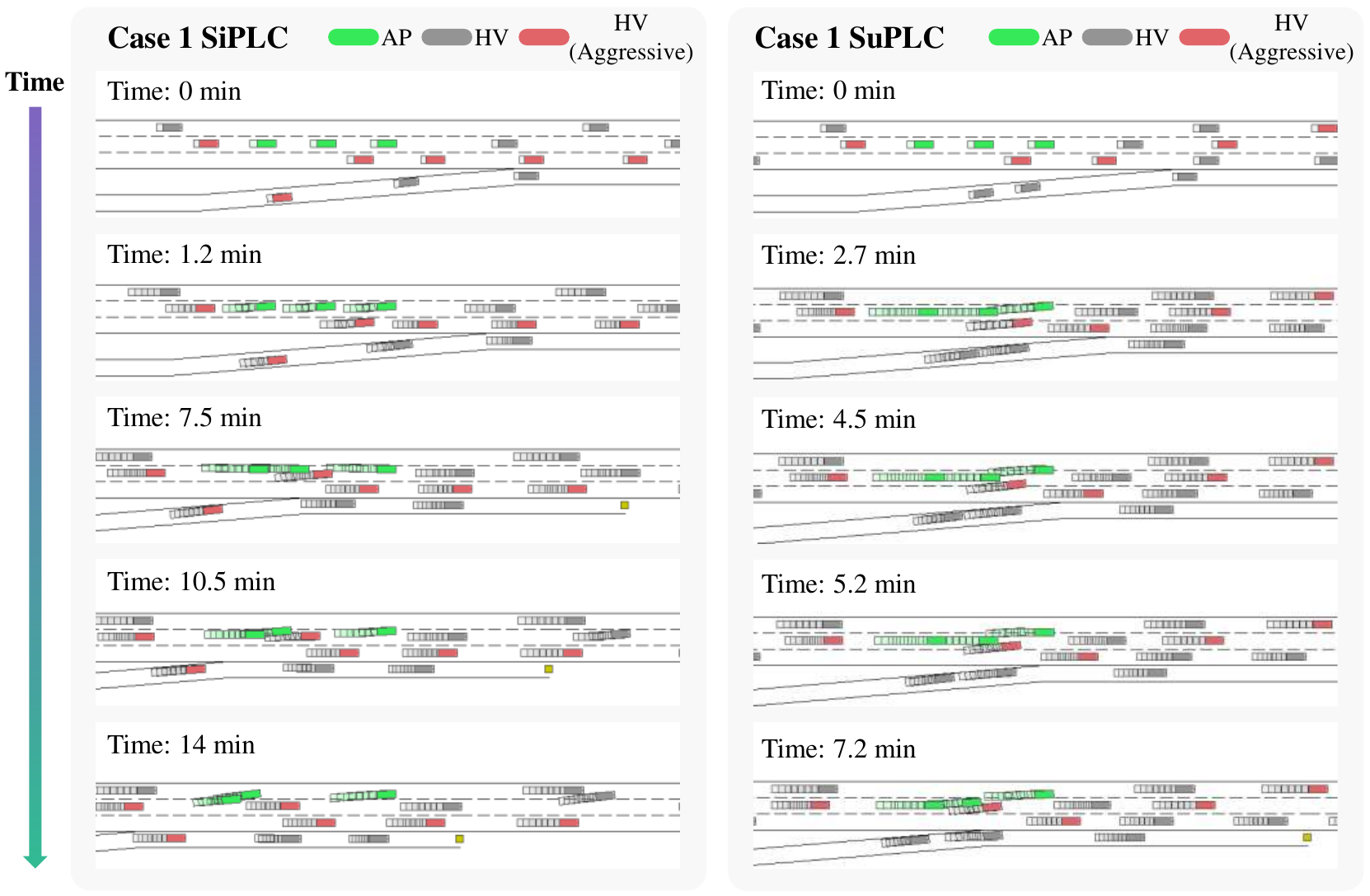}
\caption{The scenario slice diagram of SiPLC and SuPLC}
\label{fig: baseline slice}
\end{figure*}

In Case 1, we simulated a typical lateral high-risk scenario in a three-lane highway environment with ramp merging area. Lateral high risk refers to situations where the platoon’s original formation is disrupted by lateral interference from other vehicles or objects. At highway ramp merging points, unclear right-of-way between main road and ramp vehicles significantly raises the driving risk. Additionally, the continuous flow of ramp vehicles into the rightmost lane often causes congestion in that lane. To improve driving efficiency, vehicles in the rightmost lane are more likely to change lanes into the middle lane. If the controlled vehicle platoon happens to be traveling in the middle lane, the platoon's substantial spatial occupation makes it highly likely for vehicles from the rightmost lane to cut into the platoon. This increases the risk of collision within the platoon significantly.

At first, the platoon drives in tight line formation in the middle lane to decline air resistance. The distance between between the rear axle center of the lead vehicle and the rear axle center of the following vehicle is 10 m. The length of vehicle body is 5 m.

\begin{figure*}[!t]
\centering
\includegraphics[width=\textwidth]{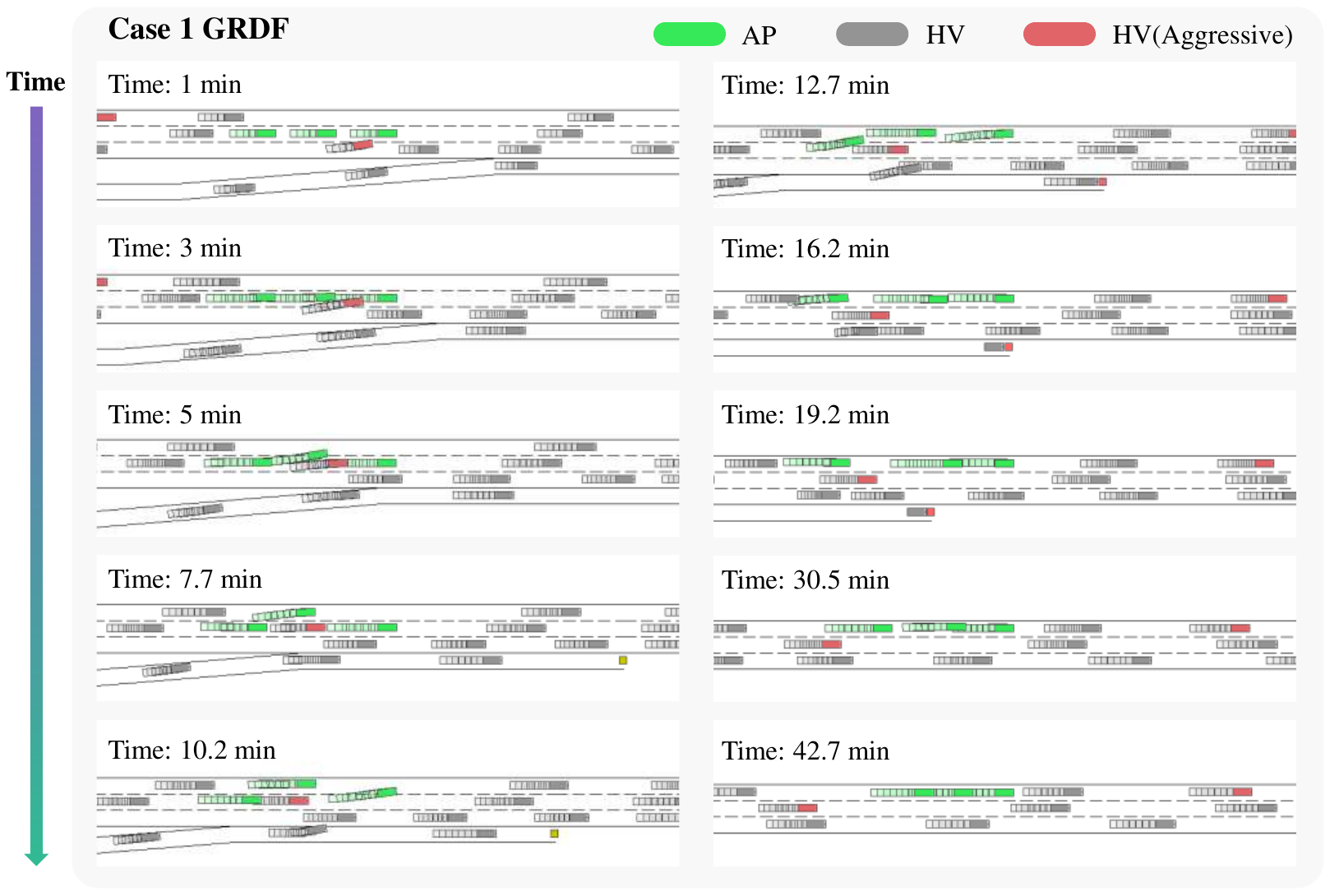}
\caption{The scenario slice diagram of GRDF}
\label{fig: GRDF slice}
\end{figure*}

To validate the advantages of the proposed multi-lane platoon reorganization strategy, we compared it with two cooperative lane-changing methods for platooning. The SiPLC proposed in 2018 \cite{r33} and SuPLC proposed in 2022 \cite{r3}, are selected as baseline models. The SiPLC allows platoon to change lanes holistically, while maintaining original formation at the same time. This will lead to conservative lane change behaviors. The platoon can only change lane as the premise of sufficient longitudinal spacing, which is hard to be satisfied in real highway environment, especially under emergency conditions. Besides, the SuPLC permits platoon to take successive lane-change, cutting through the traffic like a snake \cite{r3}. This method occupies smaller road space but extends the lane changing time window, which is difficult to implement in emergency environments. 

To visually demonstrate the differences in performance across various control models, we used scenario slice diagrams to illustrate the platoon's driving behavior. Relevant video materials can be found on the website \href{github}{https://github.com/KongAAAAAAJ/GRDF}. The platoon’s performance in the same scenario after applying the GRDF, SiPLC, and SuPLC methods is shown in Figure \ref{fig: baseline slice} and \ref{fig: GRDF slice}.

Figure \ref{fig: baseline slice} illustrates scenario slice diagrams of platoon driving under the baseline methods. The left side of Figure \ref{fig: baseline slice} shows platoon's driving behavior under SiPLC. At t=1.2 min, the platoon initiated a left lane-changing maneuver, with all vehicles in the platoon changing lanes simultaneously. As the number of merging ramp vehicles in the rightmost lane increased, a human-driven vehicle in the rear of the rightmost lane exhibited an intention to change lanes to the left. At t=7.5 min, this vehicle cut into the space between the first and second vehicles of the autonomous platoon. The second vehicle of the platoon performed an emergency brake to avoid a collision. However, the third vehicle failed to decelerate in time, resulting in a rear-end collision with the second vehicle. 

The right side of Figure \ref{fig: baseline slice} shows platoon's driving behavior under SuPLC. At t=2.7 min, the platoon initiated a left lane-changing maneuver, while the last vehicle in the rightmost lane also exhibited an intention to change lanes to the left. Since the platoon adopts a serpentine lane-changing method, the lane-changing times for the platoon’s vehicles gradually increase from the first vehicle to the last. At t=7.2 min, a human-driven vehicle cut into the space between the first and second vehicles of the platoon. By this time, the first vehicle had already completed the lane change, while the second vehicle was still in the process of changing lanes. Due to the human-driven vehicle's cut-in, the second vehicle failed to avoid the collision in time. 

Figure \ref{fig: GRDF slice} shows the vehicle platoon’s driving behaviors after applying the proposed GRDF method. As seen in Figure \ref{fig: GRDF slice}, at t=3 min, a human-driven vehicle cuts into the platoon. At t=5 min, the platoon detects the cut-in behavior of another vehicle, and the second vehicle, most affected by the cut-in, initiates a left lane change to avoid a collision. After the second vehicle changes lanes, between t=10 min and t=16 min, the remaining vehicles in the platoon sequentially change lanes with the aim of restoring the original platoon formation, ultimately successfully completing the platoon reorganization process.

This demonstrates that the SiPLC and SuPLC methods cannot effectively handle the vehicle cut-in risk during cooperative lane-change process of the platoon. These methods struggle to address unexpected dangerous scenarios, exhibiting poor robustness and insufficient safety. However, the GRDF method can autonomously organize the vehicles to complete the platoon reorganization process in an orderly manner when facing the risk of cut-in by other vehicles, effectively preventing collisions and enhancing the safety and robustness of platoon driving.

To evaluate the performance of the proposed GRDF method in terms of driving efficiency and safety, we conducted 100 simulation experiments under random traffic flow, focusing on the lateral risk scenarios. The evaluation metrics selected include collision rate, average platoon speed, average inter-vehicle distance within the platoon, and the minimum TTC of the platoon vehicles. The results of these metrics are shown in Table \ref{table case1}.

\begin{table}
\begin{center}
\caption{Performance Evaluation of Case 1}
\label{table case1}
\begin{tabular}{c c c c c}
\toprule
Index & SiPLC & SuPLC & GRDF \\
\midrule
Collision Rate & 0.884 & 0.965 & \textbf{0.021} \\
Average Speed (m/s) & 24.47 & 24.08 & \textbf{25.42} \\
Minimum TTC (s) & 3.21 & 3.39 & \textbf{7.23} \\
Average Distance (m) & \textbf{7.39} & 9.57 & 9.09 \\
\bottomrule
\end{tabular}
\end{center}
\end{table}

As shown in Table \ref{table case1}, after applying the SiPLC and SuPLC methods, the platoon experiences a relatively high collision rate in the cut-in scenarios, with values of 0.884 and 0.965, respectively. However, after applying the GRDF method, the platoon’s collision rate drops to 0.021, representing a reduction of 97.6$\%$ and 97.8$\%$, significantly enhancing collision safety. From the Minimum TTC metric, after applying the baseline models, the Minimum TTC values are 3.21 s and 3.39 s, respectively, while the TTC value for GRDF is 7.23 s, which is 2.25 times and 2.13 times higher than the baseline models. The improvement in Minimum TTC indirectly reflects the enhanced safety of the platoon’s driving. In addition, the average speed of the platoon increased from 24.47 m/s and 24.08 m/s to 25.42 m/s, representing an improvement of 3.9$\%$ and 5.6$\%$, respectively. 

\begin{figure}[!t]
\centering
\includegraphics[width=0.4\textwidth]{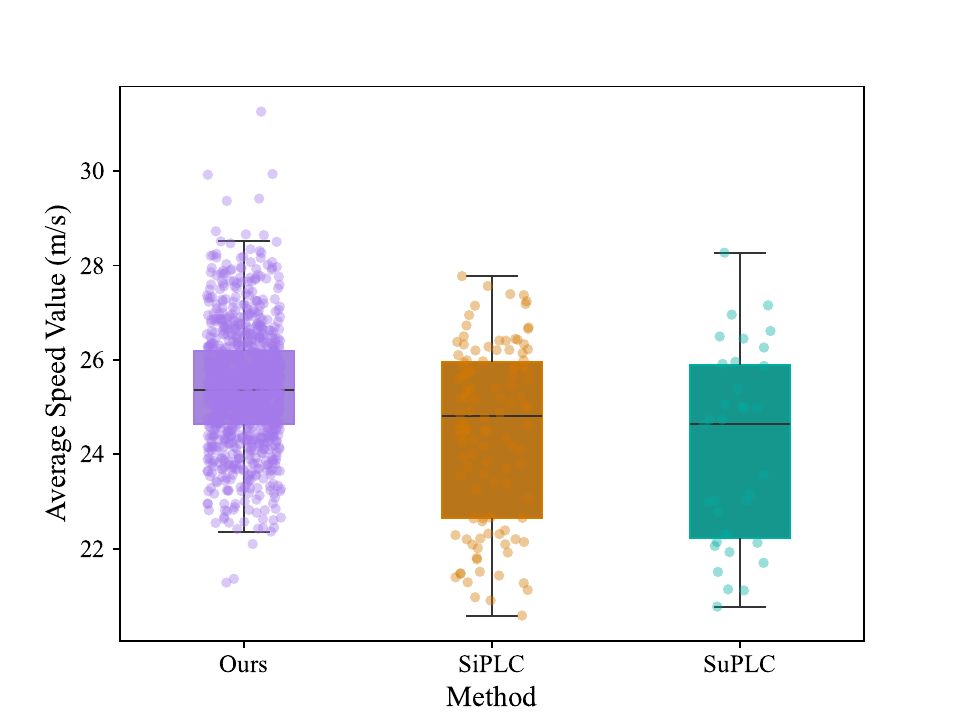}
\caption{The box plot of average speed}
\label{fig:3}
\end{figure}

\begin{figure}[!t]
\centering
\includegraphics[width=0.4\textwidth]{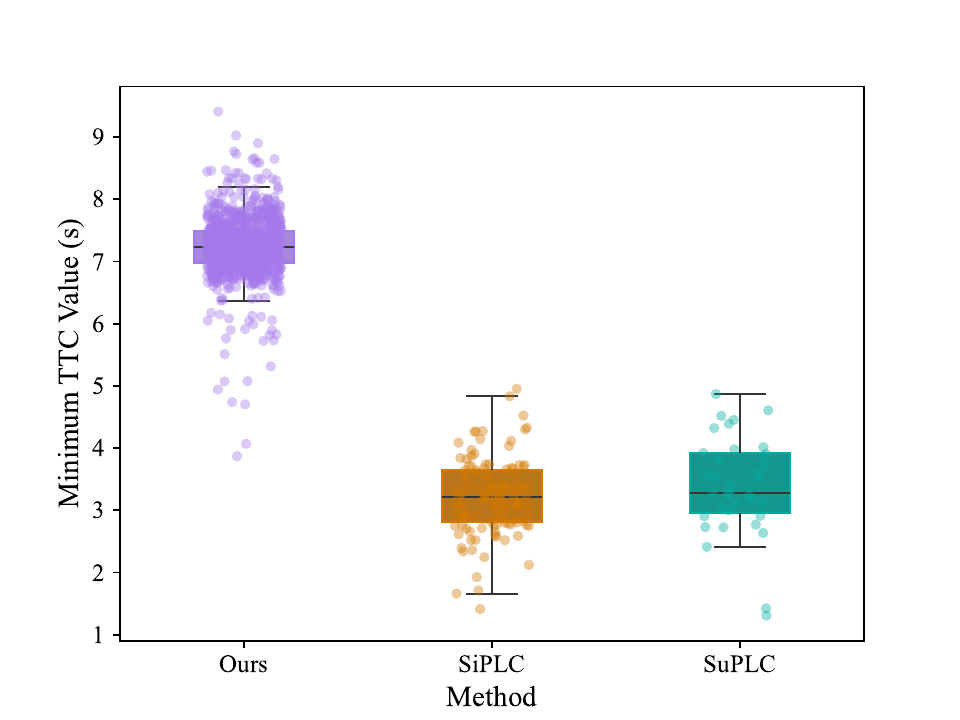}
\caption{The box plot of minimum TTC}
\label{fig:4}
\end{figure}

The data distribution of metrics are illustrated in box plots, as shown in Figure \ref{fig:3} and Figure \ref{fig:4}. The Average Speed and Minimum TTC values achieved by the GRDF model both surpass those of the baseline models. GRDF not only enhanced the safety of the platoon but also slightly improved its driving efficiency. This indicates that after applying GRDF, the platoon made better decisions, significantly improving safety without sacrificing efficiency, and effectively addressing lateral high-risk scenarios such as vehicle cut-ins.

The Average Distance metric refers to the average following distance between vehicles within the platoon, with an ideal following distance of 10 meters. SiPLC causes the vehicles to change lanes simultaneously, resulting in the shortest Average Distance of 7.39 m. SuPLC, which has the vehicles change lanes in a serpentine manner, leads to the longest Average Distance of 9.57 m. During the platoon reorganization process, the following distance inevitably increases. However, after applying the GRDF strategy, the Average Distance is 9.09 m, which is still close to the ideal following distance. This indicates that, in most cases, the platoon maintains an ideal following distance and formation, implementing the reorganization strategy only when the risk level is higher.

\subsubsection{Case 2: Longitudinal Driving Risk}
In Case 2, we simulated a typical longitudinal risk scenario for platooning in a three-lane highway environment. The longitudinal risk in this case is brought about by the rapid deceleration behavior from the vehicle in front of the whole platoon. When the front vehicle makes a sharp deceleration, the longitudinal collision risk for rear platoon will be greatly increased. The platoon needs to avoid the obstacle vehicle and reduce the loss of traffic efficiency as much as possible.

To validate the superiority of the proposed graph-theory-and-game-based formation reorganization method (GRDF-GT), we compared GRDF-GT, GRDF, and a reinforcement-learning-and-rule-based formation reorganization method (RRL), closely observing changes in relevant performance metrics.

To eliminate uncertainties, the simulation experiments were repeated 100 times under random traffic flow conditions. The results were then subjected to statistical analysis, and the key performance metrics are summarized in Table \ref{table case2}.

\begin{table}
\begin{center}
\caption{Performance Evaluation of Case 2}
\label{table case2}
\begin{tabular}{c c c c c}
\toprule
Index & RRL & GRDF & GRDF-GT \\
\midrule
Collision Rate & 0.21 & 0.01 & \textbf{0} \\
Average Speed (m/s) & 24.27 & \textbf{26.08} & 24.91 \\ 
Minimum TTC (s) & \textbf{6.50} & 4.13 & 4.08 \\
Average Distance (m) & 12.58 & 10.19 & \textbf{9.74} \\
Formation Success Rate & 0.80 & 0.88 & \textbf{0.94} \\
Formation Time (min) & 6.45 & 6.16 & \textbf{4.56} \\
\bottomrule
\end{tabular}
\end{center}
\end{table}

From Table \ref{table case2}, it can be observed that both GRDF and GRDF-GT significantly reduce the collision rate, from 0.21 to 0.01 and 0. Although the Minimum TTC indicator slightly decreases from 6.50 s to 4.13 s and 4.08 s, it remains within a safe range. Overall, the application of GRDF and GRDF-GT substantially enhances the safety of the platoon's driving performance. The Average Speed indicator also increases from 24.27 m/s to 26.08 m/s and 24.91 m/s, respectively, indicating an improvement in platoon driving efficiency while ensuring safety. 

The indicators Average Distance, Formation Success Rate, and Formation Time reflect the capabilities of the platoon during the reorganization process.

\romannumeral1) Formation Success Rate refers to whether the platoon successfully restores its original formation and completes the reorganization process within 15 min. A higher value for this indicator signifies stronger capability in executing formation reorganization.

\romannumeral2) Formation Time measures the time required for the platoon to restore its original formation from the initiation of the reorganization process. A lower value for this indicator reflects higher efficiency in the platoon's reorganization process.

As shown in Table \ref{table case2}, the Average Distance decreased from 12.58 m and 10.09 m to 9.74 m with the application of GRDF-GT. This indicates that during the reorganization process, the platoon maintained a smaller inter-vehicle distance, enhancing the overall driving efficiency. Furthermore, the Formation Success Rate was highest with GRDF-GT at 0.94, followed by GRDF at 0.88, and RRL at 0.80. In addition, the Formation Time was minimized with GRDF-GT, achieving 4.56 min, significantly lower than 6.45 min for RRL and 6.16 min for GRDF. These results demonstrate that GRDF-GT effectively enhances the platoon's formation transformation capabilities and reorganization efficiency, improving its adaptability and robustness in risk-prone scenarios.

\begin{figure}[!t]
\centering
\includegraphics[width=0.4\textwidth]{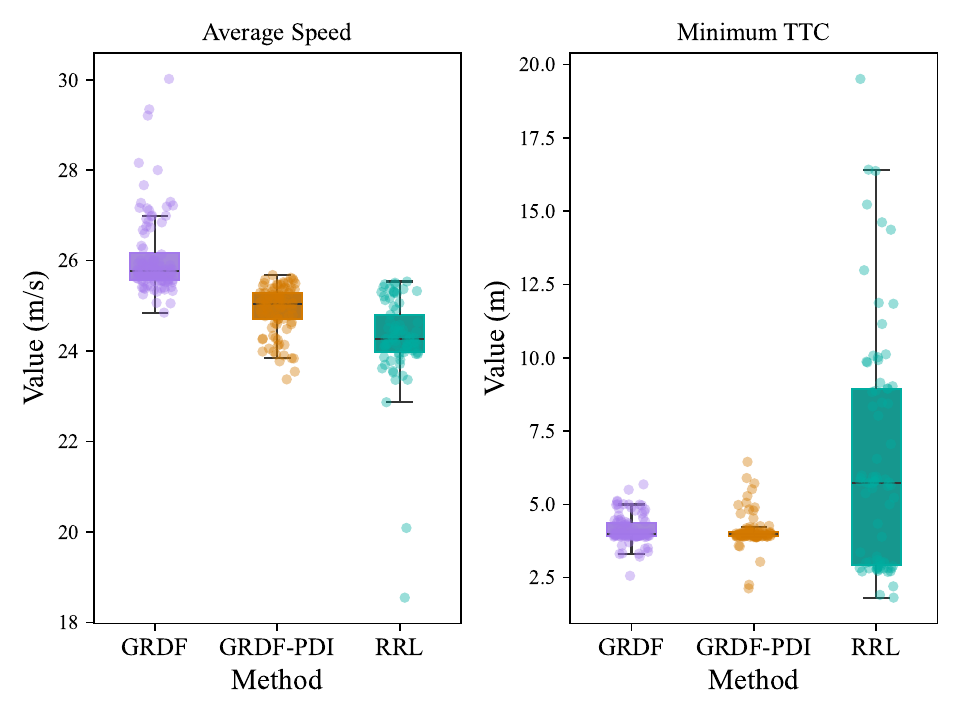}
\caption{The box plot of average speed and minimum TTC}
\label{fig:5}
\end{figure}

Meanwhile, the distribution of the Average Speed and Minimum TTC indicators is presented in the form of box plots in Figure \ref{fig:5}. It can be seen that after applying the GRDF method, the data points for the Average Speed indicator of the vehicle platoon are mostly located above those of the RRL method, demonstrating a significant advantage in terms of efficiency. In contrast, the distribution of the Minimum TTC indicator for the RRL method is quite wide, indicating that the decision-making safety of RRL is unstable across different traffic flow scenarios. There are even many data points below 2.5 s, further decreasing safety. However, after applying the GRDF and GRDF-GT models, the distribution of the platoon’s Minimum TTC indicator is more concentrated, with the vast majority of data points above 3.0 s, indicating a higher level of decision-making safety.

\begin{figure}[!t]
\centering
\includegraphics[width=0.4\textwidth]{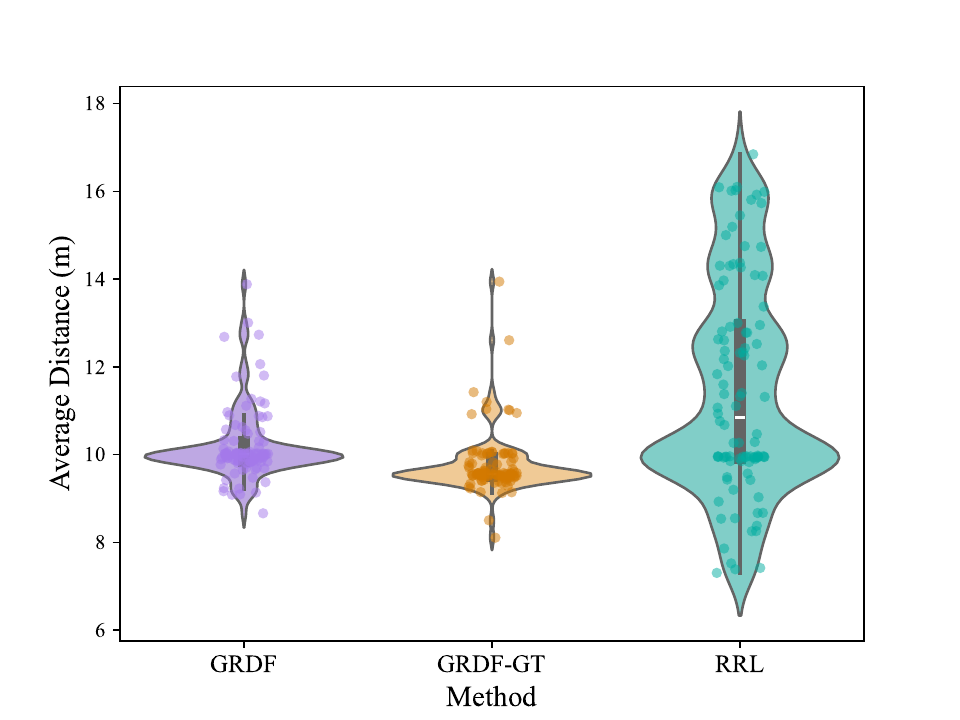}
\caption{The violin plot of average distance}
\label{fig:6}
\end{figure}

\begin{figure}[!t]
\centering
\includegraphics[width=0.4\textwidth]{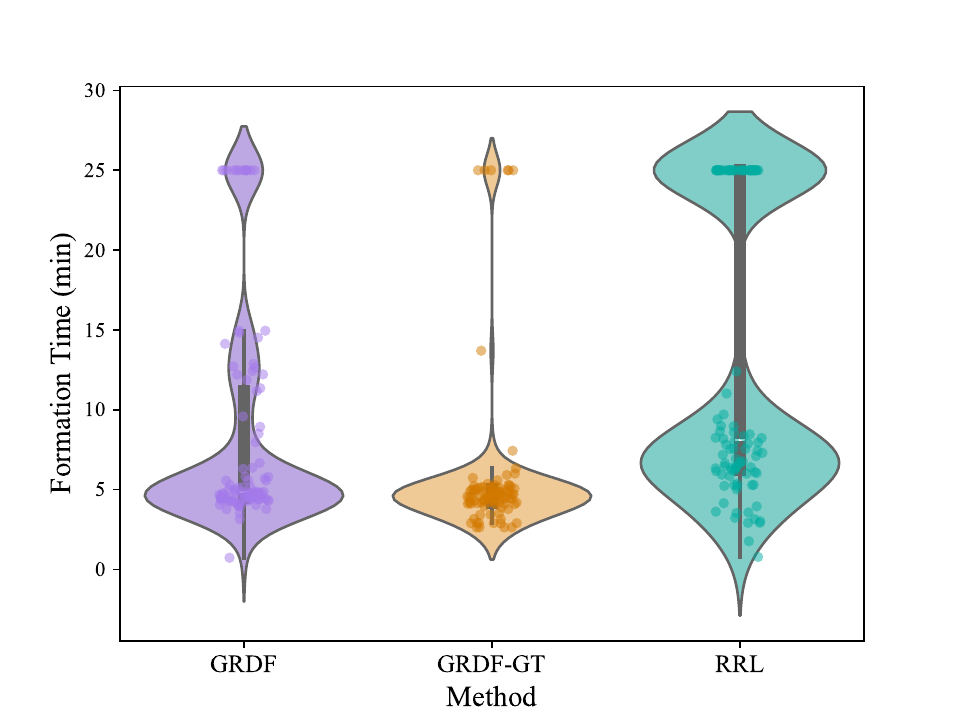}
\caption{The violin plot of formation time}
\label{fig:7}
\end{figure}

To provide a detailed presentation of the distribution density of various indicators during the platoon reorganization process, violin plots were used to analyze the Average Distance and Formation Time indicators. From Figure \ref{fig:6}, it can be observed that the Average Distance indicator for the RRL model is approximately uniformly distributed between 7 and 18 m. This suggests that the model is quite sensitive to changes in scene parameters, and the platoon maintains a relatively tight formation only in certain simpler scenarios. In contrast, both the GRDF and GRDF-GT models exhibit a more concentrated distribution of the Average Distance indicator, indicating that their decision-making methods have stronger adaptability to varying traffic scenarios. From Figure \ref{fig:7}, it can be observed that the Formation Time data for the RRL model has a wide distribution range, strongly correlating with changes in scene parameters. A significant portion of the data is distributed at the upper part of the graph, around t = 25 min, indicating that in many scenarios, the platoon does not complete the reorganization process within the specified time. In contrast, the Formation Time data for the GRDF-GT model is more concentrated and shows lower values, highlighting that the GRDF-GT model significantly improves platoon reorganization efficiency and enhances its adaptability to varying environments.

\section{Conclusion}
This paper proposes a decision-making and control framework for multi-lane platoon formation reorganization, aiming to enhance the platoon's ability to respond to dynamic emergency environment. The framework divides the reorganization process into the upper platoon-level distribution layer and the lower vehicle-level collaborative decision-making layer. In the upper layer, a reinforcement learning-based platoon formation decision algorithm is introduced, which uses a risk potential field to establish explicit risk evaluation metrics. A reward function tailored to the platoon reorganization process is designed to guide exploratory actions. In the lower layer, a coalition game algorithm is used to model cooperative relationships among vehicles. An innovative PDI is proposed to accelerate the platoon reorganization process and improve platoon driving efficiency.
In addition, the lateral and longitudinal high-risk scenarios are designed based on the directional distribution of risk sources to test the proposed method. After the testings in random traffic flows, the results show that compared to the baseline model, the proposed method reduces the collision rate in longitudinal risk scenarios from 0.884 and 0.965 to 0.021, increasing the average speed from 24.08 m/s and 24.47 m/s to 25.42 m/s. Both platoon driving safety and efficiency are improved. Besides, the proposed PDI reduces the platoon reorganization time from 6.45 min and 6.16 min to 4.56 min, significantly enhancing the efficiency of platoon reformation and configuration transitions.

\printbibliography 

@article{r1,
  title={Strategic and tactical decision-making for cooperative vehicle platooning with organized behavior on multi-lane highways},
  author={Han, Xu and Xu, Runsheng and Xia, Xin and Sathyan, Anoop and Guo, Yi and Bujanovi{\'c}, Pavle and Leslie, Ed and Goli, Mohammad and Ma, Jiaqi},
  journal={Transportation Research Part C: Emerging Technologies},
  volume={145},
  pages={103952},
  year={2022},
  publisher={Elsevier}
}

@article{r2,
  title={Collision-avoidance lane change control method for enhancing safety for connected vehicle platoon in mixed traffic environment},
  author={Ma, Yitao and Liu, Qiang and Fu, Jie and Liufu, Kangmin and Li, Qing},
  journal={Accident Analysis \& Prevention},
  volume={184},
  pages={106999},
  year={2023},
  publisher={Elsevier}
}

@article{r3,
  title={Cut through traffic like a snake: Cooperative adaptive cruise control with successive platoon lane-change capability},
  author={Wang, Haoran and Li, Xin and Zhang, Xianhong and Hu, Jia and Yan, Xuerun and Feng, Yongwei},
  journal={Journal of Intelligent Transportation Systems},
  volume={28},
  number={2},
  pages={141--162},
  year={2022},
  publisher={Taylor \& Francis}
}

@article{r4,
  title={A multistep cooperative lane change strategy for connected and autonomous vehicle platoons departing from dedicated lanes},
  author={Liu, Chenglin and Liu, Zhiguang and Xu, Zhigang and Li, Xiaopeng},
  journal={Transportation Research Part C: Emerging Technologies},
  volume={165},
  pages={104720},
  year={2024},
  publisher={Elsevier}
}

@article{r5,
  title={Make space to change lane: A cooperative adaptive cruise control lane change controller},
  author={Wang, Haoran and Lai, Jintao and Zhang, Xianhong and Zhou, Yang and Li, Shen and Hu, Jia},
  journal={Transportation research part C: emerging technologies},
  volume={143},
  pages={103847},
  year={2022},
  publisher={Elsevier}
}

@article{r6,
  title={A cooperative lane change control strategy for cooperative adaptive cruise control platoons with insufficient headway gaps},
  author={Yin, Yanduo and Gao, Zhibo and Long, Kejun and Fei, Yi},
  journal={Physica A: Statistical Mechanics and its Applications},
  volume={655},
  pages={130175},
  year={2024},
  publisher={Elsevier}
}

@article{r7,
  title={Cooperative lane-change motion planning for connected and automated vehicle platoons in multi-lane scenarios},
  author={Duan, Xuting and Sun, Chen and Tian, Daxin and Zhou, Jianshan and Cao, Dongpu},
  journal={IEEE Transactions on Intelligent Transportation Systems},
  volume={24},
  number={7},
  pages={7073--7091},
  year={2023},
  publisher={IEEE}
}

@article{r8,
  title={Decentralized Multi-Vehicle Motion Planning for Platoon Forming in Mixed Traffic Using Monte Carlo Tree Search},
  author={Liu, Chenglin and Xu, Zhigang and Liu, Zhiguang and Li, Xiaopeng and Zhang, Yuqin},
  journal={IEEE Transactions on Intelligent Transportation Systems},
  year={2024},
  publisher={IEEE}
}

@article{r9,
  title={A cooperative lane change approach for heterogeneous platoons under different communication topologies},
  author={Nie, Guangming and Xie, Bo and Lu, Huiqiu and Tian, Yantao},
  journal={IET Intelligent Transport Systems},
  volume={16},
  number={1},
  pages={53--70},
  year={2022},
  publisher={Wiley Online Library}
}

@article{r10,
  title={Coordinated lane-changing scheduling of multilane CAV platoons in heterogeneous scenarios},
  author={Liu, Qingquan and Lin, Xi and Li, Meng and Li, Li and He, Fang},
  journal={Transportation Research Part C: Emerging Technologies},
  volume={147},
  pages={103992},
  year={2023},
  publisher={Elsevier}
}

@article{r11,
  title={An Efficient Rolling-Horizon Approach for Cooperative Multi-Lane Platoon Formation With Undefined Configurations},
  author={Yang, Siwen and Xu, Yunwen and Wang, Ping and Li, Dewei},
  journal={IEEE Transactions on Intelligent Transportation Systems},
  year={2024},
  publisher={IEEE}
}

@article{r12,
  title={Hybrid MPC system for platoon based cooperative lane change control using machine learning aided distributed optimization},
  author={Zhang, Hanyu and Du, Lili and Shen, Jinglai},
  journal={Transportation Research Part B: Methodological},
  volume={159},
  pages={104--142},
  year={2022},
  publisher={Elsevier}
}

@article{r13,
  title={Cooperative platoon formation of connected and autonomous vehicles: Toward efficient merging coordination at unsignalized intersections},
  author={Deng, Zhiyun and Yang, Kaidi and Shen, Weiming and Shi, Yanjun},
  journal={IEEE Transactions on Intelligent Transportation Systems},
  volume={24},
  number={5},
  pages={5625--5639},
  year={2023},
  publisher={IEEE}
}

@article{r14,
  title={Centralized vehicle trajectory planning on general platoon sorting problem with multi-vehicle lane changing},
  author={Duan, Leyi and Wei, Yuguang and Dong, Shixin and Li, Chen},
  journal={Transportation research part C: emerging technologies},
  volume={154},
  pages={104273},
  year={2023},
  publisher={Elsevier}
}

@article{r15,
  title={Research of obstacle vehicles avoidance for automated heavy vehicle platoon by switching the formation},
  author={Kuang, Jianjie and Tan, Gangfeng and Guo, Xuexun and Pei, Xiaofei and Peng, Dengzhi},
  journal={IET Intelligent Transport Systems},
  volume={18},
  number={4},
  pages={630--644},
  year={2024},
  publisher={Wiley Online Library}
}

@article{r16,
  title={Deep Q-network based multi-layer safety lane changing strategy for vehicle platoon},
  author={Zhang, Jinqi and Yan, Maode and Zuo, Lei},
  journal={IET Intelligent Transport Systems},
  volume={18},
  number={4},
  pages={645--656},
  year={2024},
  publisher={Wiley Online Library}
}

@article{r17,
  title={Platoon-aware cooperative lane-changing strategy for connected automated vehicles in mixed traffic flow},
  author={Jiang, Yangsheng and Tan, Li and Xiao, Guosheng and Wu, Yunxia and Yao, Zhihong},
  journal={Physica A: Statistical Mechanics and its Applications},
  volume={640},
  pages={129689},
  year={2024},
  publisher={Elsevier}
}

@article{r18,
  title={A Multi-agent Reinforcement Learning Based Control Method for Connected and Autonomous Vehicles in A Mixed Platoon},
  author={Xu, Yaqi and Shi, Yan and Tong, Xiaolu and Chen, Shanzhi and Ge, Yuming},
  journal={IEEE Transactions on Vehicular Technology},
  year={2024},
  publisher={IEEE}
}

@article{r19,
  title={A reinforcement learning-based vehicle platoon control strategy for reducing energy consumption in traffic oscillations},
  author={Li, Meng and Cao, Zehong and Li, Zhibin},
  journal={IEEE Transactions on Neural Networks and Learning Systems},
  volume={32},
  number={12},
  pages={5309--5322},
  year={2021},
  publisher={IEEE}
}

@ARTICLE{r20,
  title={Collaborative Control of Vehicle Platoon Based on Deep Reinforcement Learning}, 
  author={Chen, Jianzhong and Wu, Xiaobao and Lv, Zekai and Xu, Zhihe and Wang, Wenjie},
  journal={IEEE Transactions on Vehicular Technology}, 
  volume={73},
  number={10},
  pages={14399-14414},
  year={2024},
  publisher={IEEE}
}

@article{r21,
  title={Enforcing Cooperative Safety for Reinforcement Learning-based Mixed-Autonomy Platoon Control},
  author={Zhou, Jingyuan and Yan, Longhao and Liang, Jinhao and Yang, Kaidi},
  journal={arXiv preprint arXiv:2411.10031},
  year={2024}
}

@article{r22,
  title={A coevolutionary algorithm for cooperative platoon formation of connected and automated vehicles},
  author={Deng, Zhiyun and Fan, Jiaxin and Shi, Yanjun and Shen, Weiming},
  journal={IEEE Transactions on Vehicular Technology},
  volume={71},
  number={12},
  pages={12461--12474},
  year={2022},
  publisher={IEEE}
}

@article{r23,
  title={Behavioral Decision-Making Approach for Vehicle Platoon Control: Two Noncooperative Game Models},
  author={Liu, Yang and Zong, Changfu and Dai, Changhua and Zheng, Hongyu and Zhang, Dong},
  journal={IEEE Transactions on Transportation Electrification},
  volume={9},
  number={3},
  pages={4626--4638},
  year={2023},
  publisher={IEEE}
}

@article{r24,
  title={Deep reinforcement learning based incentive mechanism design for platoon autonomous driving with social effect},
  author={Li, Bo and Xie, Kan and Huang, Xumin and Wu, Yuan and Xie, Shengli},
  journal={IEEE Transactions on Vehicular Technology},
  volume={71},
  number={7},
  pages={7719--7729},
  year={2022},
  publisher={IEEE}
}

@article{r25,
  title={Game theory-based decision-making and iterative predictive lateral control for cooperative obstacle avoidance of guided vehicle platoon},
  author={Gong, Xinle and Liang, Sheng and Wang, Bowen and Zhang, Wei},
  journal={IEEE Transactions on Vehicular Technology},
  volume={72},
  number={6},
  pages={7051--7066},
  year={2023},
  publisher={IEEE}
}

@article{r26,
  title={Where to Decide? Centralized Versus Distributed Vehicle Assignment for Platoon Formation},
  author={Heinovski, Julian and Dressler, Falko},
  journal={IEEE Transactions on Intelligent Transportation Systems},
  year={2024},
  publisher={IEEE}
}

@article{r27,
  title={A novel multimode hybrid control method for cooperative driving of an automated vehicle platoon},
  author={Ma, Yulin and Li, Zhixiong and Malekian, Reza and Zheng, Sifa and Sotelo, Miguel Angel},
  journal={IEEE Internet of Things Journal},
  volume={8},
  number={7},
  pages={5822--5838},
  year={2020},
  publisher={IEEE}
}

@article{r28,
  title={Digital twin empowered cooperative trajectory planning of platoon vehicles for collision avoidance with unexpected obstacles},
  author={Du, Hao and Leng, Supeng and He, Jianhua and Xiong, Kai and Zhou, Longyu},
  journal={Digital Communications and Networks},
  year={2023},
  publisher={Elsevier}
}

@article{r29,
  title={A homogeneous multi-vehicle cooperative group decision-making method in complicated mixed traffic scenarios},
  author={Wang, Yuning and Li, Jinhao and Ke, Tianqi and Ke, Zehong and Jiang, Junkai and Xu, Shaobing and Wang, Jianqiang},
  journal={Transportation Research Part C: Emerging Technologies},
  volume={167},
  pages={104833},
  year={2024},
  publisher={Elsevier}
}

@article{r30,
  title={Deep reinforcement learning for intelligent transportation systems: A survey},
  author={Haydari, Ammar and Y{\i}lmaz, Yasin},
  journal={IEEE Transactions on Intelligent Transportation Systems},
  volume={23},
  number={1},
  pages={11--32},
  year={2020},
  publisher={IEEE}
}

@article{r31,
  title={Concept, principle and modeling of driving risk field based on driver-vehicle-road interaction},
  author={Wang, Jian-qiang and Wu, Jian and Li, Yang},
  journal={China Journal of Highway and Transport},
  volume={29},
  number={1},
  pages={105--114},
  year={2016}
}

@article{r32,
  title={Coalitional game theory for communication networks},
  author={Saad, Walid and Han, Zhu and Debbah, M{\'e}rouane and Hjorungnes, Are and Basar, Tamer},
  journal={Ieee signal processing magazine},
  volume={26},
  number={5},
  pages={77--97},
  year={2009},
  publisher={IEEE}
}

@article{r33,
  title={Kinematic design for platoon-lane-change maneuvers},
  author={Hsu, Harry Chia-Hung and Liu, Alan},
  journal={IEEE Transactions on Intelligent Transportation Systems},
  volume={9},
  number={1},
  pages={185--190},
  year={2008},
  publisher={IEEE}
}

@article{r35,
  title={European Research Project’s Contributions to a Safer Automated Road Traffic},
  author={Fahrenkrog, Felix and Reithinger, Susanne and G{\"u}lsen, Burak and Raisch, Florian},
  journal={Automotive Innovation},
  volume={6},
  number={4},
  pages={521--530},
  year={2023},
  publisher={Springer}
}

@article{pan2023research,
  title={Research on multi-lane energy-saving driving strategy of connected electric vehicle based on vehicle speed prediction},
  author={Pan, Chaofeng and Li, Yuan and Wang, Jian and Liang, Jun and Jinyama, Ho},
  journal={Green Energy and Intelligent Transportation},
  volume={2},
  number={6},
  pages={100127},
  year={2023},
  publisher={Elsevier}
}

@article{li2023continual,
  title={Continual driver behaviour learning for connected vehicles and intelligent transportation systems: Framework, survey and challenges},
  author={Li, Zirui and Gong, Cheng and Lin, Yunlong and Li, Guopeng and Wang, Xinwei and Lu, Chao and Wang, Miao and Chen, Shanzhi and Gong, Jianwei},
  journal={Green Energy and Intelligent Transportation},
  pages={100103},
  year={2023},
  publisher={Elsevier}
}

@article{naeem2023optimal,
  title={Optimal-Control-Based Eco-Driving Solution for Connected Battery Electric Vehicle on a Signalized Route},
  author={Naeem, Hafiz Muhammad Yasir and Butt, Yasir Awais and Ahmed, Qadeer and Bhatti, Aamer Iqbal},
  journal={Automotive Innovation},
  volume={6},
  number={4},
  pages={586--596},
  year={2023},
  publisher={Springer}
}

@article{xu2024towards,
  title={Towards Safe and Robust Autonomous Vehicle Platooning: A Self-Organizing Cooperative Control Framework},
  author={Xu, Chengkai and Deng, Zihao and Liu, Jiaqi and Huang, Chao and Hang, Peng},
  journal={arXiv preprint arXiv:2408.09468},
  year={2024}
}

\end{document}